\PassOptionsToPackage{unicode}{hyperref}
\PassOptionsToPackage{hyphens}{url}
\documentclass[11pt,a4paper]{article}
\usepackage{amsmath,amssymb}
\usepackage[margin=1in]{geometry}
\usepackage{pdflscape}
\usepackage{iftex}
\ifPDFTeX
  \usepackage[T1]{fontenc}
  \usepackage[utf8]{inputenc}
  \usepackage{textcomp} 
\else 
  \usepackage{unicode-math} 
  \defaultfontfeatures{Scale=MatchLowercase}
  \defaultfontfeatures[\rmfamily]{Ligatures=TeX,Scale=1}
\fi
\usepackage{lmodern}
\DeclareUnicodeCharacter{221E}{\ensuremath{\infty}}
\DeclareUnicodeCharacter{2212}{\ensuremath{-}}
\DeclareUnicodeCharacter{03B1}{\ensuremath{\alpha}}
\DeclareUnicodeCharacter{03B2}{\ensuremath{\beta}}
\DeclareUnicodeCharacter{03C6}{\ensuremath{\phi}}
\ifPDFTeX\else
\fi
\IfFileExists{upquote.sty}{\usepackage{upquote}}{}
\IfFileExists{microtype.sty}{
  \usepackage[]{microtype}
  \UseMicrotypeSet[protrusion]{basicmath} 
}{}
\makeatletter
\@ifundefined{KOMAClassName}{
  \IfFileExists{parskip.sty}{%
    \usepackage{parskip}
  }{
    \setlength{\parindent}{0pt}
    \setlength{\parskip}{6pt plus 2pt minus 1pt}}
}{
  \KOMAoptions{parskip=half}}
\makeatother
\usepackage{xcolor}
\usepackage{longtable,booktabs,array}
\usepackage{multirow}
\usepackage{calc} 
\usepackage{etoolbox}
\makeatletter
\patchcmd\longtable{\par}{\if@noskipsec\mbox{}\fi\par}{}{}
\makeatother
\IfFileExists{footnotehyper.sty}{\usepackage{footnotehyper}}{\usepackage{footnote}}
\makesavenoteenv{longtable}
\usepackage{graphicx}
\makeatletter
\def\maxwidth{\ifdim\Gin@nat@width>\linewidth\linewidth\else\Gin@nat@width\fi}
\def\maxheight{\ifdim\Gin@nat@height>\textheight\textheight\else\Gin@nat@height\fi}
\makeatother
\setkeys{Gin}{width=\maxwidth,height=\maxheight,keepaspectratio}
\makeatletter
\def\fps@figure{htbp}
\makeatother
\ifLuaTeX
  \usepackage{selnolig}  
\fi
\usepackage{bookmark}
\IfFileExists{xurl.sty}{\usepackage{xurl}}{} 
\hypersetup{
  hidelinks,
  pdfcreator={LaTeX reconstructed from an original author Word manuscript},
  pdfsubject={Author manuscript; published DOI 10.1016/j.nucengdes.2024.113082},
  pdfkeywords={IEA-R1 research reactor,research reactor core compaction,thermal neutron flux enhancement,lutetium-177 (Lu-177) production,medical radioisotope production,SCALE KENO-VI criticality calculations,Monte Carlo neutron transport,neutron irradiation positions,radioisotope irradiation yield,core reactivity analysis}}

\author{}
\date{}

\begin{document}
\noindent\fbox{\begin{minipage}{0.95\linewidth}
\small\textbf{Author manuscript -- publication record.} \textit{An earlier coauthor-reviewed Word version, December 2023; not asserted to be the finally accepted R2 revision.}\\
The definitive peer-reviewed paper appeared in \emph{Nuclear Engineering and Design 421 (2024), 113082}.\\
Original DOI: \href{https://doi.org/10.1016/j.nucengdes.2024.113082}{10.1016/j.nucengdes.2024.113082}. \textbf{Please cite the published article as the version of record.}
\end{minipage}}\par\medskip

\textbf{Core reduction for increasing neutron flux and radioisotope production in the IEA-R1 research reactor}

Felipe Viggiano\textsuperscript{a,*}, Giovanni L. Stefani\textsuperscript{b}, Frederico A. Genezini\textsuperscript{a}, João M. L. Moreira\textsuperscript{c,}, Caio J. C. M. Cunha\textsuperscript{d}.

\emph{\textsuperscript{a}} \emph{Instituto de Pesquisas Energéticas e Nucleares (IPEN / CNEN - SP)}

\emph{05508-000, São Paulo, SP, Brazil}

\emph{felipeviggiano@gmail.com, fredzini@ipen.br}

\emph{\textsuperscript{b}} \emph{Universidade Federal do Rio de Janeiro (UFRJ)}

\emph{Departamento de Engenharia Nuclear -- Escola Politécnica,}

\emph{21941-909, Rio de Janeiro, RJ, Brazil}

\emph{laranjogiovanni@poli.ufrj.br}

\emph{\textsuperscript{c}} \emph{Universidade Federal do ABC (UFABC)}

\emph{Programa de Pós-graduação em Energia}

\emph{09210-580, Santo André, SP, Brazil}

\emph{joao.moreira@ufabc.edu.br}

\emph{\textsuperscript{d}} \emph{Universidade Federal de Pernambuco (UFPE)}

\emph{Departamento de Energia Nuclear}

\emph{50740-545, Recife, PE, Brazil}

\href{mailto:cjcmrcunha@gmail.com}{\emph{cjcmrcunha@gmail.com}}

\emph{* corresponding author}

\textbf{ABSTRACT}

This work aims to present and analyze a new reactor core configuration for the IEA-R1 reactor core with the objective of enhancing its neutron irradiation capabilities. The reactor was modeled using the KENO-VI module from the SCALE Code System, provided by ORNL. Results have shown that implementing a near-cylindrical configuration with 20 fuel elements in the IEA-R1 core could substantially increase neutron flux, particularly in the center irradiation positions such as the EIBe (Beryllium Irradiated Element). After comparing various configurations, we conducted a brief analysis of the proposed core configuration's potential to produce \textsuperscript{177}Lu, resulting in a 34\% increase in the specific activity of \textsuperscript{177}Lu considering a fuel power cycle around 21 days.

Keywords: \textsuperscript{177}Lu production, Monte Carlo method, KENO-VI, neutron flux, IEA-R1 reactor

\section{INTRODUCTION}\label{introduction}

The increasing demand for radioisotopes for medical and industrial applications have induced reactor operators to study alternatives to increase their facilities capacity to produce radioisotopes and provide new services {[}1,4-6{]}. The alternatives for improving radioisotope production in research reactors include the utilization of high-density fuel {[}4{]}, placing special irradiator elements in the core to increase local epithermal and thermal neutron flux {[}5,6,7,13{]}, placing more efficient reflectors {[}4{]}, and changing the core geometry {[}6{]}. In Brazil the demand for radioisotopes is growing and places significant challenges to radioisotope producers {[}2,3{]} including to the operators of the IEA-R1 research reactor, the oldest nuclear reactor in Brazil, built under the Atoms for Peace Program by Babcok \& Wilcox Company and achieved its first criticality in 1957 {[}1{]}. It is currently the main supplier of radiochemical including all the \textsuperscript{99m}Tc production {[}4{]}.

One of these demands is the production of \textsuperscript{177}Lu for use in treatments based on target radionuclide therapy and other medical applications. Research reactors are interesting for producing this radionuclide as long as they can reach thermal neutron fluxes greater than 0.8x10\textsuperscript{14} n cm\textsuperscript{-2}s\textsuperscript{-1} and can operate continuously for periods between 14 and 21 days {[}13-15{]}. To address the increasing demand studies have been carried out to enhance the IEA-R1 reactor radioisotope production. Stefani et al. {[}7{]} investigated new core configurations with the Monte Carlo SERPENT code {[}5{]} aiming to elevate the neutron flux at the reactor's irradiation positions. They concluded that a near cylindrical configuration could significantly increase the neutron flux in irradiation positions.

Recently, the IEA-R1 was modelled using the SCALE 6.2 {[}11{]} and previous studies are being reviewed to verify the consistency of this new model. In this study, the work from Laranjo et. al. was conducted considering this new simulation model, in addition to the insertion of the burn up rates in the fuel elements.

In the Methods section we present the information employed to generate a 3D model as close as possible to the real-life reactor, and some of the techniques applied to obtain the needed information from the simulation's modules. In Validation section we have compared the output data from the standard configuration with data obtained from the actual core, such as xenon build-up, reactivity, control rod influence and thermal flux profile. In the Results section, the different configurations were compared with each other, first all of them, and further analysis were conducted only with the promising ones. New configurations, beyond the analyzed ones in the reference study, were modeled to verify the full potential of the core configuration changing, such as new irradiation positions and \textsuperscript{177}Lu production.

\section{METHODS AND DATA}\label{methods-and-data}

The approach taken in this study involved proposing seven different configurations for the IEA-R1 reactor core, with the primary goal of enhancing the neutron flux levels in the irradiation positions. These positions were identified as: EIF, EIBRA 1 EIBRA 2, EIBe A and EIBe B. To achieve this enhancement, the approach taken was to change fuel elements configuration in the core, changing the reflector material from Graphite to Beryllium, and changing the position of irradiator positions in the core. The reactor core's thermal power level remained at 4.5 MW.

The fuel and moderator temperature coefficients of reactivity were calculated since significant changes in the core were implemented. Furthermore, the fuel cycle length was also verified to account for potential alterations resulting from the removal of fuel element in specific configurations.

To validate the accuracy of k\textsubscript{eff} and neutron flux estimations at the irradiation positions we compared the calculated results for the E01 configuration with those presented in the IEA-R1 research reactor Safety Analysis Report. Below we present the reactor core descriptions, details of the seven core configurations and specifics from the KENO-VI calculations.

Finally, a steady state single heated channel thermal analysis was performed in order to verify the thermal behavior of the chosen configuration.

\subsection{\texorpdfstring{Neutron flux, k\textsubscript{eff} and burnup calculations }{Neutron flux, keff and burnup calculations }}\label{neutron-flux-keff-and-burnup-calculations}

The reactor core was modeled based on configuration number 263, which was in place during the year 2022, using the KENO-VI code, a Monte Carlo module from the SCALE Code System. This method was chosen due to its reduced computational requirements in comparison with deterministic approaches when dealing with high complexity modelling scenarios such as nuclear reactors. All the simulations were executed using the 252 multigroup ENDF VII.0 library, embedded in the SCALE code system. The multigroup library was chosen due to its good performance capabilities {[}10{]}. This work focused on the thermal flux, considering neutrons with energies below 0.625 eV.

To achieve accurate flux, simulations were conducted using the mesh feature of KENO-VI, which allows to compute the fluxes in a grid-based system. In this work the grid was defined with each mesh having dimensions 2.59 cm x 2.7 cm x 1.36 cm, returning in a flux map with 211,680 positions.

For depletion (burnup) and activation simulations the TRITON module of SCALE was used, providing a depletion sequence using the same model employed in KENO-VI {[}11{]}. The construct the depleted IEA-R1 configuration E01 (configuration 263) each fuel element burnup was updated using the Burnup Monitoring Report provided by IEA-R1 operator {[}9{]}.

\textbf{2.2. Reactor core description}

The IEA-R1 reactor is a 5 MW Material Test reactor (MTR) located at the Institute for Nuclear and Energy Research (IPEN) at the University of São Paulo campus, São Paulo, Brazil. The IEA-R1 reactor core consists of a matrix plate with 80-positions (8x10) accommodating fuel elements, neutron reflectors and irradiation positions. The MTR type fuel elements have 19.75 \% enriched uranium, the dispersed uranium silicide alloy (U\textsubscript{3}Si\textsubscript{2}-Al) cladded with an Aluminum plate.

Figure 1 shows the core configuration 263 of the IEA-R1 research reactor, comprising 24 Standard Fuel Elements (SFE) and 4 Control Fuel Elements (CFE) surrounded by beryllium and graphite reflectors. For material irradiation, 11 positions are available throughout the core {[}1{]}.

\includegraphics[width=5.38542in,height=3.22569in]{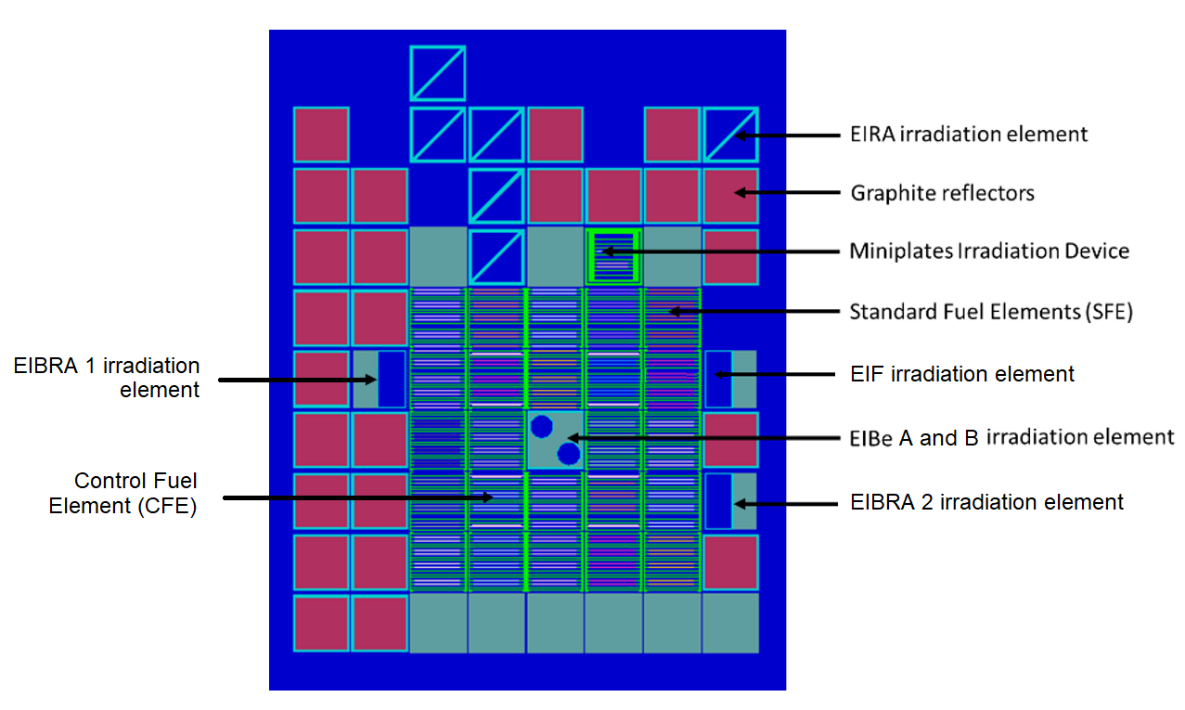}

\emph{Figure 1: Core configuration number 263 (Configuration E01) from the IEA-R1 research reactor. Top view identifying fuel elements, irradiation positions, neutron reflectors and other devices.}

Each SFE comprises 18 fuel plates measuring 7.17 cm x 0.152 cm x 62.5 cm. The fuel meat is positioned within a framing module and compressed between two aluminum plates, which act as cladding. The active region itself has dimensions 6 cm x 0.076 cm x 60 cm.

The CFE share similarities with the SFE, albeit with twelve fuel plates. The remaining spaces serve as a pathway for the insertion of control rods. These control rods are made of Ag-In-Cd alloy with proportion of 80\%, 15\% and 5\%, respectively. Figure 2 shows the modeled SFE and CFE used in this work, being blue, green, and white the water (moderator), aluminum (cladding and structure), and control rods, respectively. The colored plates situated within the plates represent the active region of the fuel.

\begin{longtable}[]{@{}
  >{\raggedright\arraybackslash}p{(\columnwidth - 2\tabcolsep) * \real{0.5079}}
  >{\raggedright\arraybackslash}p{(\columnwidth - 2\tabcolsep) * \real{0.4921}}@{}}
\toprule\noalign{}
\endhead
\bottomrule\noalign{}
\endlastfoot
\includegraphics[width=1.50278in,height=1.575in]{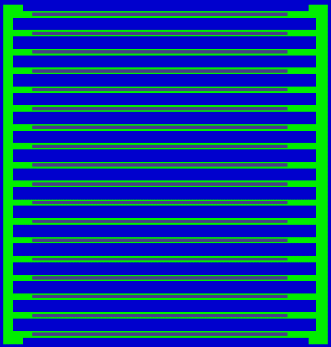}

SFE & \includegraphics[width=1.50278in,height=1.575in]{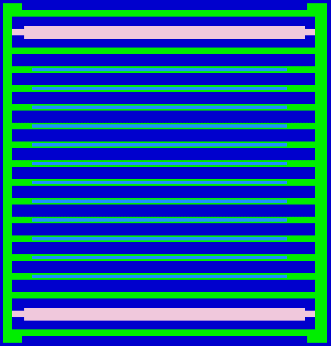}

CFE \\
\end{longtable}

\emph{Figure 2: Top view of Standard Fuel Element (SFE) and Control Fuel Element (CFE).}

\begin{enumerate}
\def\labelenumi{\arabic{enumi}.}
\item
  \textbf{Proposed core configurations to enhance neutron flux at irradiation positions}
\end{enumerate}

In addition to the standard configuration (E01), six alternative arrangements were modeled and designated as configurations E02 to E07 as shown in Figure 3. In the configuration E02, the Graphite reflectors adjacent to the fuel elements were replaced by Beryllium ones. In the E03, all the graphite reflectors present in the core were replaced by Beryllium reflectors. In the E04 configuration, four of the standard fuel elements were removed in a cross pattern, and in the E05 configuration, an ``X'' pattern was followed to remove four standard fuel elements.

The removal of 4 SFE elements from the core opens the opportunity to incorporate four additional irradiation positions. Two additional configurations (E06 and E07) were devised to represent these new conditions. In configuration E06, EIRA elements were inserted on the positions where the SFE were removed, and in configuration E07, EIBRA elements were inserted for this purpose.

\begin{longtable}[]{@{}
  >{\raggedright\arraybackslash}p{(\columnwidth - 2\tabcolsep) * \real{0.4999}}
  >{\raggedright\arraybackslash}p{(\columnwidth - 2\tabcolsep) * \real{0.5001}}@{}}
\toprule\noalign{}
\endhead
\bottomrule\noalign{}
\endlastfoot
\includegraphics[width=1.50833in,height=1.96667in]{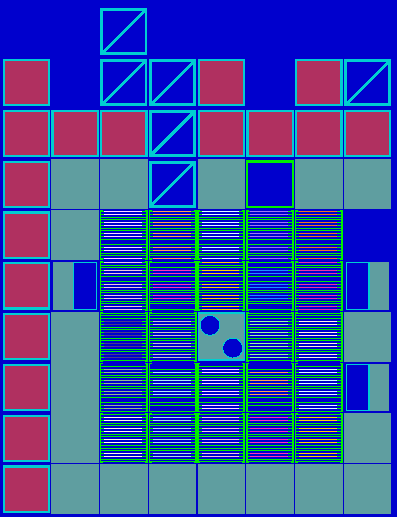}

Configuration E02 with 24 fuel elements & \includegraphics[width=1.525in,height=1.96667in]{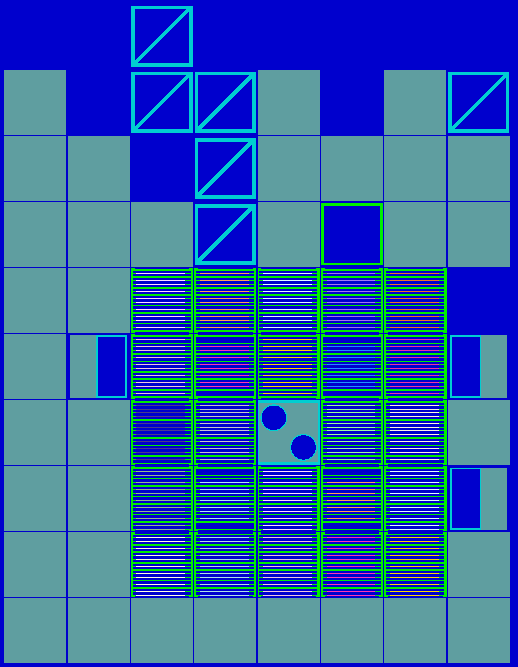}

Configuration E03 with 24 fuel elements \\
\includegraphics[width=1.51667in,height=1.96667in]{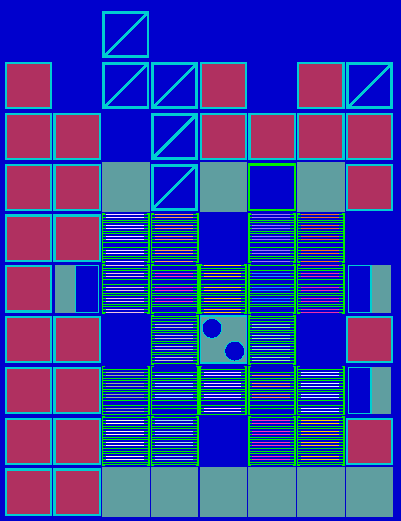}

Configuration E04 with 20 fuel elements & \includegraphics[width=1.525in,height=1.96667in]{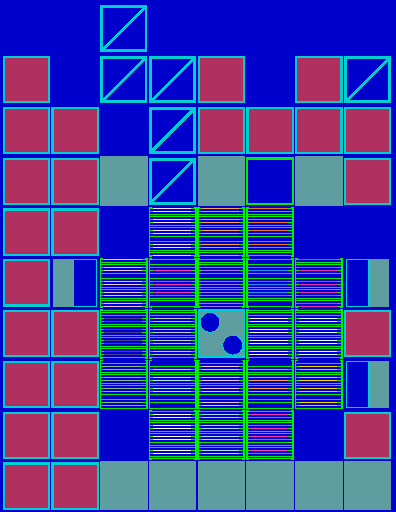}

Configuration E05 with 20 fuel elements \\
\includegraphics[width=1.49931in,height=1.96875in]{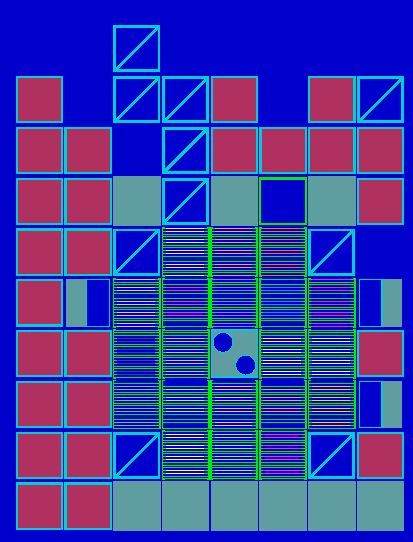}

Configuration E06 with 20 fuel elements & \includegraphics[width=1.52153in,height=1.96875in]{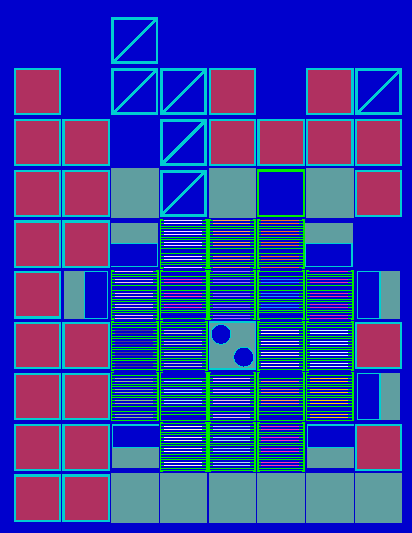}

Configuration E07 with 20 fuel elements \\
\end{longtable}

\emph{Figure 3: Configurations E02 to E07.}

A single-channel model was developed using the ANSYS CFX computational fluid dynamics (CFD) package to analyze the thermal characteristics of this new core proposal for the IEA-R1 reactor. We used the benchmark {[}16{]} as a reference for defining the geometric model, boundary, and operating conditions of the IEA-R1 standard core and the new proposal, respectively. We designed a 1/2 three-dimensional model of a fuel element (composed of the fuel plate, cladding, and coolant) with the ANSYS SpaceClaim tool, taking advantage of the symmetry conditions, as illustrated in Figure 4.

\includegraphics[width=5.90486in,height=2.84722in]{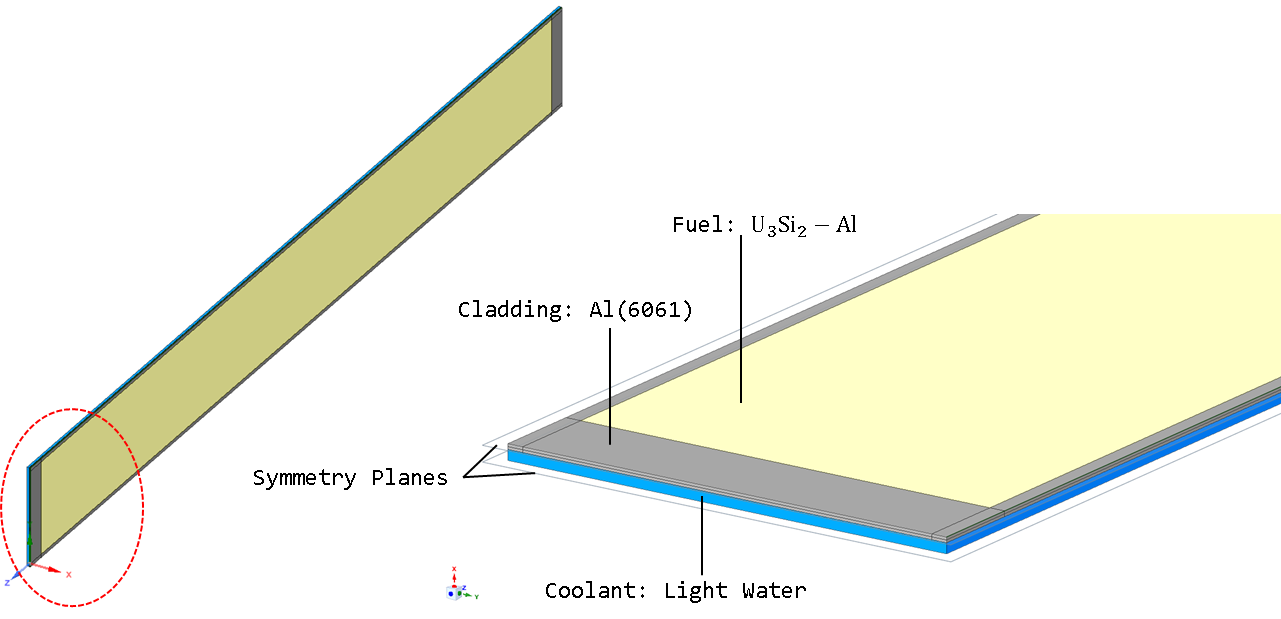}

\emph{Figure 4: Fuel element isometric view.}

The proposed new core holds the structural dimensions of the standard core with fewer fuel assemblies. Therefore, the geometry of the fuel element was modeled according to the dimensions presented in Table 1.

Table 1: Fuel element dimensions {[}16{]}.

\begin{longtable}[]{@{}
  >{\raggedright\arraybackslash}p{(\columnwidth - 6\tabcolsep) * \real{0.4428}}
  >{\raggedright\arraybackslash}p{(\columnwidth - 6\tabcolsep) * \real{0.2286}}
  >{\raggedright\arraybackslash}p{(\columnwidth - 6\tabcolsep) * \real{0.1547}}
  >{\raggedright\arraybackslash}p{(\columnwidth - 6\tabcolsep) * \real{0.1738}}@{}}
\toprule\noalign{}
\endhead
\bottomrule\noalign{}
\endlastfoot
\multirow{2}{=}{\textbf{Domain}} & \multicolumn{3}{>{\raggedright\arraybackslash}p{(\columnwidth - 6\tabcolsep) * \real{0.5572} + 4\tabcolsep}@{}}{%
\textbf{Values (mm)}} \\
& \textbf{Thickness} & \textbf{Width} & \textbf{Length} \\
\textbf{Fuel} & 0.76 & 62.6 & 600 \\
\textbf{Cladding} & 1.52 & 67.1 & 625 \\
\textbf{Coolant} & 2.89 & 67.1 & 625 \\
\end{longtable}

A spatial discretization of the computational domain was designed using the ANSYS Meshing tool, in which we built a mesh of hexahedrons for all regions under analysis, taking advantage of the geometry sweep condition. Figure 5 illustrates the mesh used in the present study.

\includegraphics[width=4.70556in,height=3.30764in]{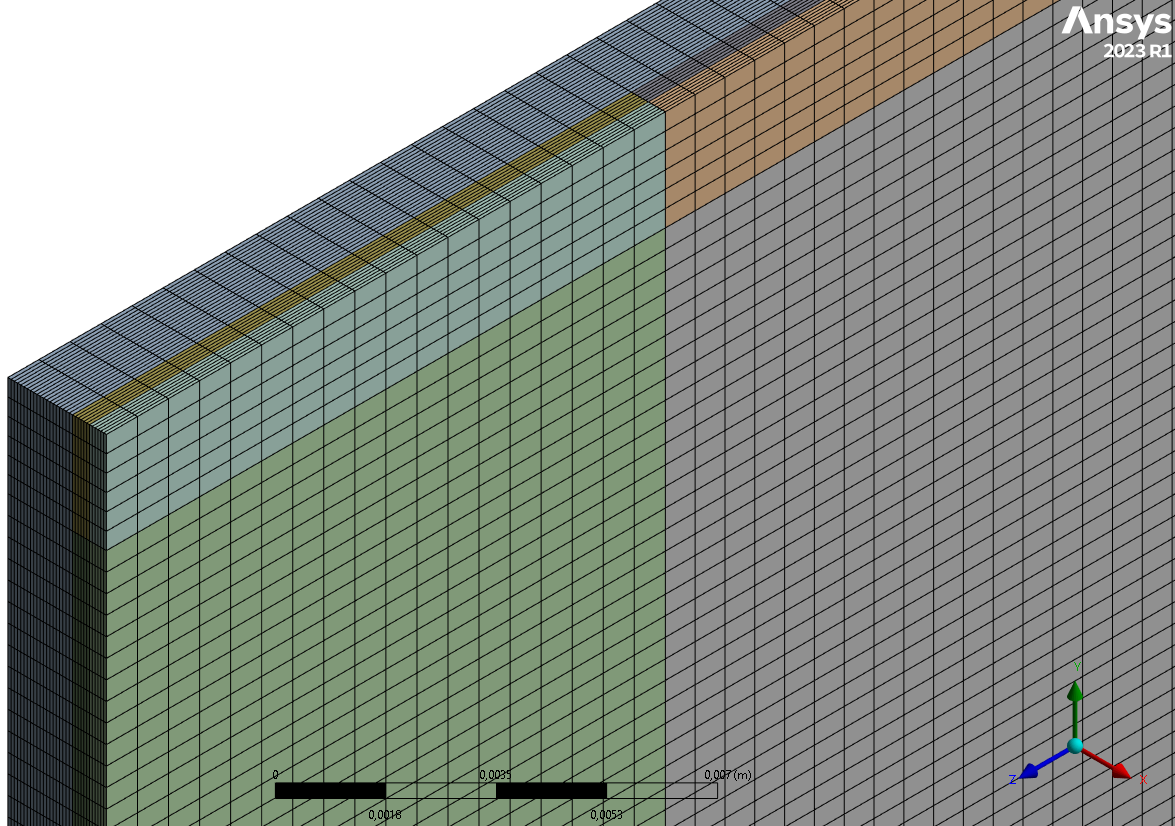}

\emph{Figure 5: Isometric view of the fuel element mesh.}

The fuel and cladding regions were discretized with six elements in thickness, while the coolant region has 20 elements to capture local phenomena accurately. As a result, the mesh used in this study has 5.04 million nodes and 4.85 million hexahedron elements. Table 2 presents the mesh quality parameters and their respective results.

\emph{Table 2: Mesh metrics.}

\begin{longtable}[]{@{}
  >{\raggedright\arraybackslash}p{(\columnwidth - 8\tabcolsep) * \real{0.2553}}
  >{\raggedright\arraybackslash}p{(\columnwidth - 8\tabcolsep) * \real{0.1853}}
  >{\raggedright\arraybackslash}p{(\columnwidth - 8\tabcolsep) * \real{0.1854}}
  >{\raggedright\arraybackslash}p{(\columnwidth - 8\tabcolsep) * \real{0.1852}}
  >{\raggedright\arraybackslash}p{(\columnwidth - 8\tabcolsep) * \real{0.1888}}@{}}
\toprule\noalign{}
\endhead
\bottomrule\noalign{}
\endlastfoot
\textbf{Parameter} & \textbf{Minimum} & \textbf{Maximum} & \textbf{Average} & \textbf{Standard deviation} \\
\textbf{Skewness} & 1.3057e-010 & 5.9914e-004 & 8.3443e-006 & 3.3438e-005 \\
\textbf{Orthogonal Quality} & 1 & 1 & 1 & 2.136e-007 \\
\textbf{Aspect Ratio} & 9.2266 & 10.965 & 9.7299 & 0.63459 \\
\end{longtable}

By taking advantage of the geometry\textquotesingle s sweep behavior, a mesh of excellent quality was build, which reflects the response accurately and with an optimized calculation time. To achieve this scenario, in addition to the refined work with the mesh stage, we performed physical modeling of the problem based on the benchmark {[}16{]} and the paper {[}17{]} to define the boundary/operation conditions and the thermophysical properties of materials, respectively. Table 3 presents the boundary conditions for steady-state analysis of the standard IEA-R1 reactor core.

\emph{Table 3: IEA-R1 steady state boundary conditions {[}16{]}.}

\begin{longtable}[]{@{}
  >{\raggedright\arraybackslash}p{(\columnwidth - 2\tabcolsep) * \real{0.6848}}
  >{\raggedright\arraybackslash}p{(\columnwidth - 2\tabcolsep) * \real{0.3152}}@{}}
\toprule\noalign{}
\endhead
\bottomrule\noalign{}
\endlastfoot
\textbf{Parameter} & \textbf{Value} \\
\textbf{Nominal System Pressure} & 1.7 bar \\
\textbf{Nominal Reactor Power} & 3.5 -- 5.0 MW \\
\textbf{Mass Flow Rate (per Instrumented Fuel Assembly)} & 6.27 kg/s \\
\textbf{Reactor Core Inlet temperature} & 31.61 -- 33.43 \ensuremath{^\circ}C \\
\end{longtable}

In the current study, the materials used, except water, are not defined in ANSYS CFX. As a result, we set the thermophysical properties of the fuel (U3Si2 -- Al) and the cladding (AL -- 6061). Table 4 lists the constant values for thermal conductivity, density, and specific heat capacity suggested by {[}17{]}.

\emph{Table 4: Materials thermophysical properties {[}17{]}.}

\begin{longtable}[]{@{}
  >{\raggedright\arraybackslash}p{(\columnwidth - 4\tabcolsep) * \real{0.2356}}
  >{\raggedright\arraybackslash}p{(\columnwidth - 4\tabcolsep) * \real{0.4653}}
  >{\raggedright\arraybackslash}p{(\columnwidth - 4\tabcolsep) * \real{0.2991}}@{}}
\toprule\noalign{}
\endhead
\bottomrule\noalign{}
\endlastfoot
\textbf{Domain} & \textbf{Parameter} & \textbf{Value} \\
\multirow{3}{=}{\textbf{Fuel}

\textbf{(U\textsubscript{3}Si\textsubscript{2} -- Al)}} & \textbf{Thermal Conductivity} & 50.03 W/mK \\
& \textbf{Density} & 6030 kg/m³ \\
& \textbf{Specific Heat Capacity} & 338.4 J/kgK \\
\multirow{3}{=}{\textbf{Cladding}

\textbf{(AL -- 6061)}} & \textbf{Thermal Conductivity} & 176.01 W/mK \\
& \textbf{Density} & 2700 kg/m³ \\
& \textbf{Specific Heat Capacity} & 998.56 J/kgK \\
\end{longtable}

The IAPWS-IF97 library was used to get water properties, and a the standard k-epsilon turbulence model with a scalable wall function was set. A high-resolution option was assumed for the advective scheme, with a convergence criterion of 1E-6 and a maximum number of interactions of 1E3.

\subsection{\texorpdfstring{Production of \textsuperscript{177}Lu}{Production of 177Lu}}\label{production-of-177lu}

The literature informs that viable 177Lu production for target radionuclide therapy and other medical applications occur with thermal neutron flux levels greater than 0.8x10\textsuperscript{14} n cm\textsuperscript{-2}s\textsuperscript{-1}. The final activity of \textsuperscript{177}Lu is proportional to the thermal neutron flux at the irradiation location and the irradiation period {[}12-15{]}. To calculate specific \textsuperscript{177}Lu activities we considered 28 days of full-power operation.

To assess potential of producing \textsuperscript{177}Lu, a sample was placed in the position with the highest flux (1.21x10\textsuperscript{14} at \textasciitilde13 cm below the center of the active region). This work has considered only the direct route of production \textsuperscript{176}Lu(n,\ensuremath{\gamma})\textsuperscript{177}Lu due to the lower logistics costs {[}12{]}.

The aluminum containment filled with air, as used in experimental situations, was modeled to hold the sample. The sample itself consists of a 100 \% \textsuperscript{176}Lu cuboid with a volume of 0.5 cm³ (dimensions 1 cm x 1 cm x 0.5 cm). Special care was taken to account for the self-shielding effect on highly enriched lutetium samples {[}13{]}. Figure 6 shows the modelled containment with the \textsuperscript{176}Lu sample inside.

\includegraphics[width=5.28403in,height=2.58611in]{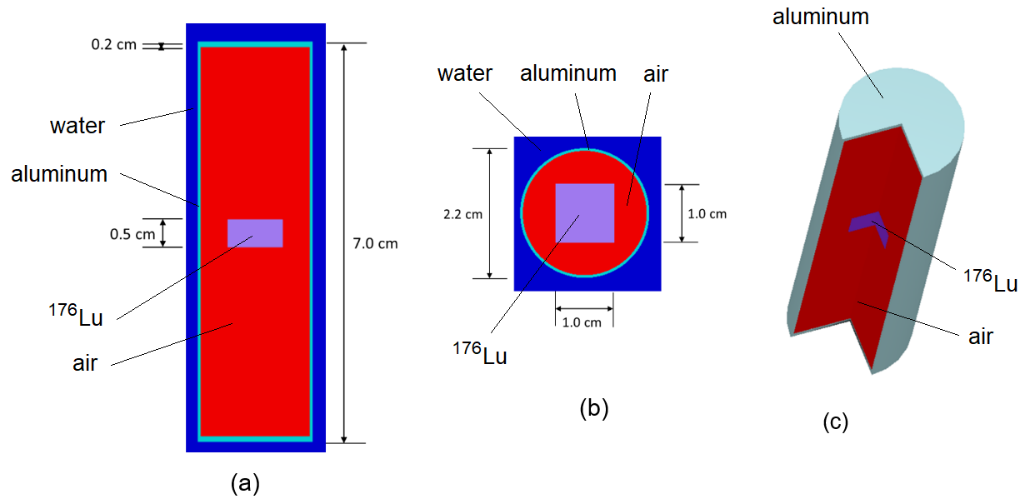}

\emph{Figure 6: Aluminum containment. Front view (a), Top view (b) and 3D view (excluded the water surrounding).}

\section{RESULTS AND DISCUSSIONS}\label{results-and-discussions}

We start presenting the validation of the neutronic calculations and follow presenting results regarding safety and operational parameters of the IEA-R1 reactor for the proposed new configurations. In the final subsection we present the impact on the 177Lu in the IEA-R1 reactor. At each section we discuss the obtained results.

\section{Validation}\label{validation}

Once modelled some of the key parameters from the core configuration 263 were compared benchmark results. The parameters were the effective multiplication factor, xenon buildup and the core thermal flux at mapped positions inside different irradiator elements. The benchmark values were taken from the IEA-R1 reactor Safety Analysis Report (SAR) or furnished by the IEA-R1 operator.

The xenon buildup in the reactor core was obtained from a 60-hour simulation (48 hours at 100\% power followed by 12 hours at 0\% power). was conducted. The model yielded an average \textsuperscript{135}Xe equilibrium concentration of 2.29 \ensuremath{\mu}g/cm³ resulting in a total insertion of negative reactivity of 3135 \ensuremath{\pm} 123 pcm. The SAR stated a negative reactivity insertion of 3200 pcm in equilibrium, resulting in a discrepancy of 2.07 \%.

Figure 7 presents the thermal neutron flux at several core locations used for sample irradiation. The results were compared with the thermal flux map provided by the reactor's management to the users.

\includegraphics[width=5.19028in,height=2.9125in]{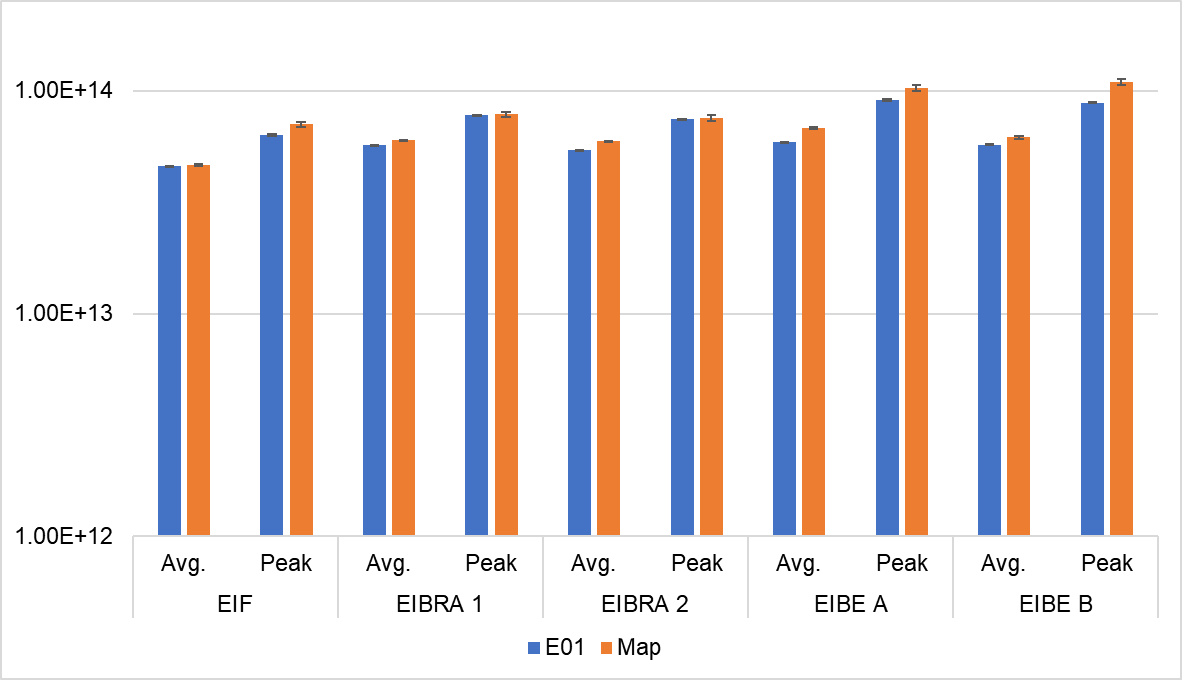}

\emph{Figure 7: Thermal neutron flux (n/cm\textsuperscript{2}s) comparison at different irradiation locations. Map means results provided by the IEA-R1 operator and E01 means results from this model. The results include average and peak thermal flux at the irradiation locations.}

The comparison indicates a good adherence of the model with the flux map provided by the reactor operator since the mapped neutron flux could present errors up to 50\% depending on the irradiation time {[}14{]}. The model presented combined discrepancies of 7.5 \% and 16.8 \% for average and peak thermal neutron fluxes, respectively.

\subsection{Thermal neutron flux}\label{thermal-neutron-flux}

To compare the different configurations (E01, E02, E03, E04 and E05), the thermal neutron flux was obtained at five irradiation positions: EIF, EIBRA 1, EIBRA 2, EIBe A and EIBe B, identified in Figure 1. Thermal neutrons were defined as those with energies below 0.625 eV. All neutron flux simulations were conducted with 100 million histories (2000 generations with 50.000 neutrons each) and the control rods adjusted to the critical position. Figures 8 to 12 show the resulted thermal flux in each position for the configurations under review. It is possible to see that a great impact is made in the EIBe positions due to its central position in the core.

\includegraphics[width=4.36667in,height=3.18056in]{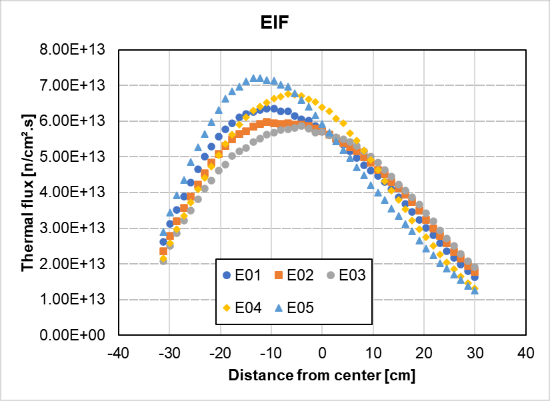}

\emph{Figure 8: Thermal neutron flux in the irradiation position EIF.}

\includegraphics[width=4.69097in,height=3.35764in]{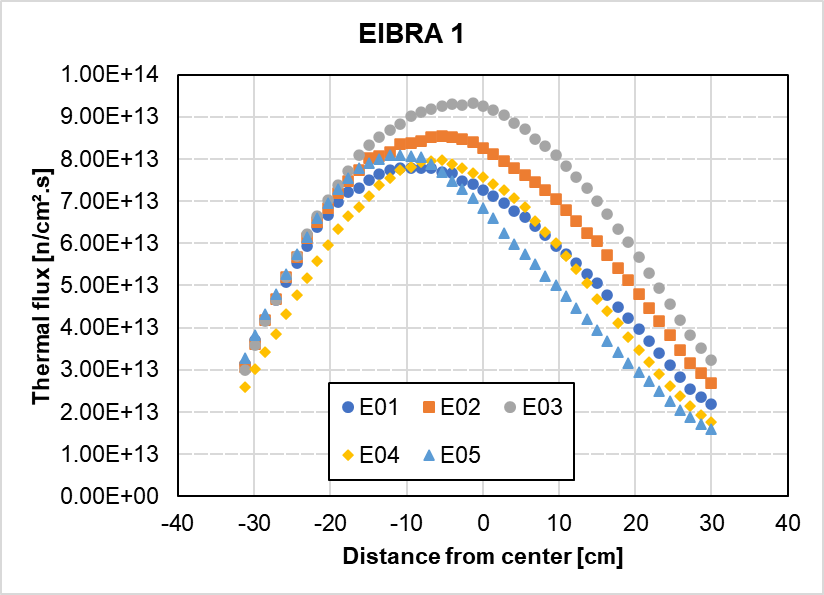}

\emph{Figure 9: Thermal neutron flux in the irradiation position EIBRA 1.}

\includegraphics[width=4.88194in,height=3.51944in]{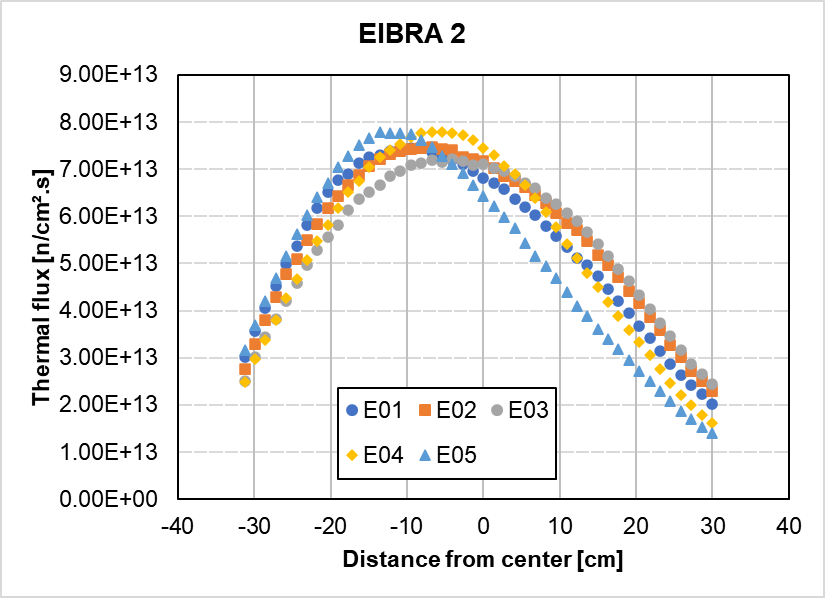}

\emph{Figure 10: Thermal neutron flux in the irradiation position EIBRA 2.}

\includegraphics[width=4.68056in,height=3.35208in]{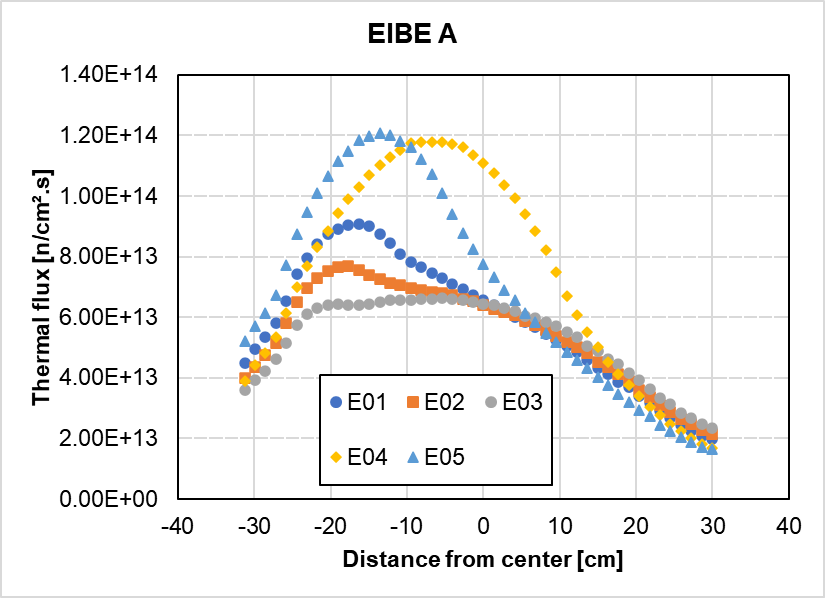}

\emph{Figure 11: Thermal neutron flux in the irradiation position EIBE A.}

\begin{longtable}[]{@{}
  >{\raggedright\arraybackslash}p{(\columnwidth - 0\tabcolsep) * \real{1.0000}}@{}}
\toprule\noalign{}
\endhead
\bottomrule\noalign{}
\endlastfoot
\includegraphics[width=4.66389in,height=3.31875in]{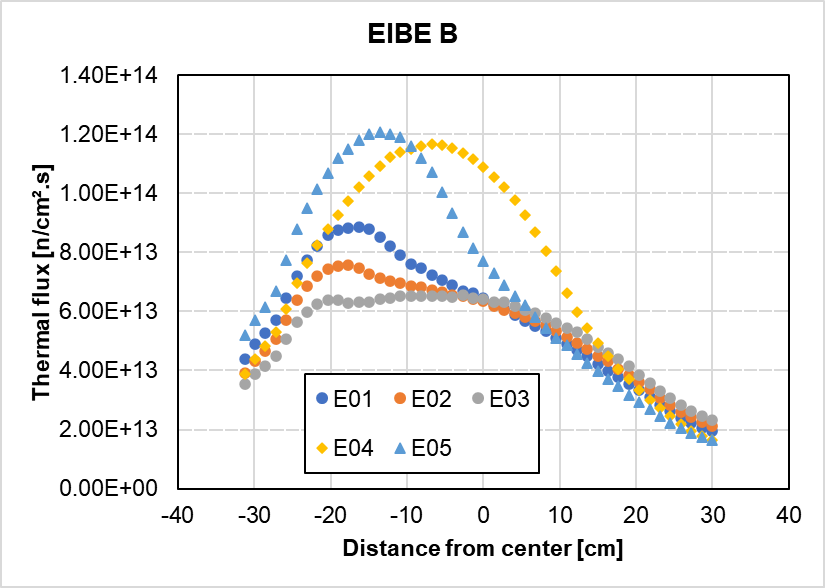}

\emph{Figure 12: Thermal neutron flux in the irradiation position EIBE B.} \\
\end{longtable}

Table 5 summarize the fluxes obtained with respect to the standard configuration E01. Configurations E04 and E05 results present greater increases in the thermal neutron fluxes. At irradiation positions at the EIBE irradiation element it reaches over 30\% increase.

\emph{Table 5: Thermal neutron flux variation with respect to the E01 configuration.}

\begin{longtable}[]{@{}
  >{\raggedright\arraybackslash}p{(\columnwidth - 12\tabcolsep) * \real{0.1578}}
  >{\raggedright\arraybackslash}p{(\columnwidth - 12\tabcolsep) * \real{0.1054}}
  >{\raggedright\arraybackslash}p{(\columnwidth - 12\tabcolsep) * \real{0.1544}}
  >{\raggedright\arraybackslash}p{(\columnwidth - 12\tabcolsep) * \real{0.1527}}
  >{\raggedright\arraybackslash}p{(\columnwidth - 12\tabcolsep) * \real{0.1432}}
  >{\raggedright\arraybackslash}p{(\columnwidth - 12\tabcolsep) * \real{0.1509}}
  >{\raggedright\arraybackslash}p{(\columnwidth - 12\tabcolsep) * \real{0.1355}}@{}}
\toprule\noalign{}
\endhead
\bottomrule\noalign{}
\endlastfoot
& & \textbf{E01} & \textbf{E02} & \textbf{E03} & \textbf{E04} & \textbf{E05} \\
\multirow{2}{=}{\textbf{EIF}} & \textbf{Avg.} & 4.59E+13 & -2.09\% & -4.18\% & -0.42\% & 2.96\% \\
& \textbf{Peak} & 6.35E+13 & -5.93\% & -7.54\% & 6.37\% & 13.49\% \\
\multirow{2}{=}{\textbf{EIBRA 1}} & \textbf{Avg.} & 5.73E+13 & 11.20\% & 22.21\% & -4.47\% & -6.01\% \\
& \textbf{Peak} & 7.79E+13 & 9.52\% & 19.67\% & 2.45\% & 3.84\% \\
\multirow{2}{=}{\textbf{EIBRA 2}} & \textbf{Avg.} & 5.44E+13 & 2.73\% & 0.34\% & -2.05\% & -5.88\% \\
& \textbf{Peak} & 7.46E+13 & 0.19\% & -3.16\% & 4.54\% & 4.38\% \\
\multirow{2}{=}{\textbf{EIBE A}} & \textbf{Avg.} & 5.89E+13 & -6.50\% & -9.64\% & 26.83\% & 18.16\% \\
& \textbf{Peak} & 9.09E+13 & -15.41\% & -27.12\% & 29.75\% & 32.69\% \\
\multirow{2}{=}{\textbf{EIBE B}} & \textbf{Avg.} & 5.74E+13 & -5.50\% & -8.38\% & 28.27\% & 20.92\% \\
& \textbf{Peak} & 8.86E+13 & -14.54\% & -26.13\% & 31.65\% & 36.46\% \\
\multirow{2}{=}{\textbf{TOTAL}} & \textbf{Avg.} & 5.48E+13 & -0.01\% & 0.18\% & 10.29\% & 6.36\% \\
& \textbf{Peak} & 9.09E+13 & -6.14\% & 2.56\% & 29.75\% & 32.93\% \\
\end{longtable}

The configurations E04 and E05 with 20 fuel elements provide the greater average and peak neutron flux increases. Configuration E02 presents thermal neutron flux reduction in the irradiation positions and E03 presents a marginal increase of less than 3 \% at peak locations.

\subsection{Moderator and fuel temperature coefficients of reactivity}\label{moderator-and-fuel-temperature-coefficients-of-reactivity}

The E05 configuration was chosen, since it provided the most significant increase in neutron flux at peak locations to the fuel and moderator temperature coefficients of reactivity. The water temperature was raised from 294.5 K to 354.5 K in three steps of 20 K each, while the fuel temperature remained constant at 356.5 K. Similarly, the fuel coefficient was examined by simulating three steps of 30 K, 50 K and 100K, ranging from 294.5 K to 474.5 K. Figure 13 shows the evolution of the multiplication (k\textsubscript{eff}) factor with the changes in the moderator and fuel temperatures.

\begin{longtable}[]{@{}
  >{\raggedright\arraybackslash}p{(\columnwidth - 2\tabcolsep) * \real{0.4989}}
  >{\raggedright\arraybackslash}p{(\columnwidth - 2\tabcolsep) * \real{0.5011}}@{}}
\toprule\noalign{}
\endhead
\bottomrule\noalign{}
\endlastfoot
\includegraphics[width=3.14167in,height=1.85069in]{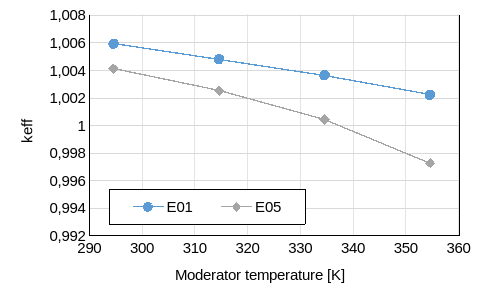}

(a) & \includegraphics[width=3.02917in,height=1.7875in]{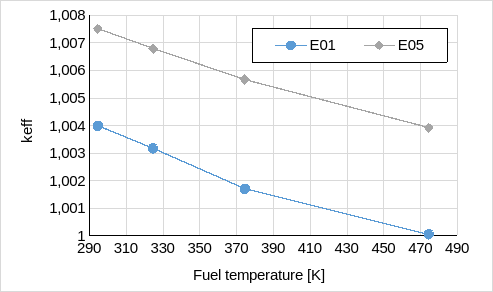}

(b) \\
\end{longtable}

\emph{Figure 13: k\textsubscript{eff} evolution for temperatures changes in moderator (a) and fuel (b).}

Table 6 shows the resulting moderator and temperature coefficients for E05 configuration in comparison with the standard one (E01).

\emph{Table 6: Fuel and moderator coefficients}

\begin{longtable}[]{@{}
  >{\raggedright\arraybackslash}p{(\columnwidth - 6\tabcolsep) * \real{0.2231}}
  >{\raggedright\arraybackslash}p{(\columnwidth - 6\tabcolsep) * \real{0.2779}}
  >{\raggedright\arraybackslash}p{(\columnwidth - 6\tabcolsep) * \real{0.3329}}
  >{\raggedright\arraybackslash}p{(\columnwidth - 6\tabcolsep) * \real{0.1661}}@{}}
\toprule\noalign{}
\endhead
\bottomrule\noalign{}
\endlastfoot
\textbf{Configuration} & \textbf{Fuel temperature (K)} & \textbf{Moderator temperature (K)} & \textbf{Value}

\textbf{(pcm/K)} \\
\multicolumn{4}{@{}>{\raggedright\arraybackslash}p{(\columnwidth - 6\tabcolsep) * \real{1.0000} + 6\tabcolsep}@{}}{%
\textbf{Moderator temperature coefficient of reactivity}} \\
\multirow{3}{=}{E01} & \multirow{3}{=}{356.5} & 314.5 & -12.00 \\
& & 334.5 & -16.15 \\
& & 354.5 & -18.98 \\
& & & \\
\multirow{3}{=}{E05} & \multirow{3}{=}{356.5} & 314.5 & -7.84 \\
& & 334.5 & -10.39 \\
& & 354.5 & -15.99 \\
& & & \\
\multicolumn{4}{@{}>{\raggedright\arraybackslash}p{(\columnwidth - 6\tabcolsep) * \real{1.0000} + 6\tabcolsep}@{}}{%
\textbf{Fuel temperature coefficient of reactivity}} \\
\multirow{3}{=}{E01} & 309.5 & \multirow{3}{=}{354.5} & -2.7 \\
& 349.5 & & -2.2 \\
& 424.5 & & -1.7 \\
& & & \\
\multirow{3}{=}{E05} & 309.5 & \multirow{3}{=}{354.5} & -2.4 \\
& 349.5 & & -2.2 \\
& 424.5 & & -1.7 \\
\end{longtable}

\subsection{\texorpdfstring{Fuel cycle length }{Fuel cycle length }}\label{fuel-cycle-length}

The nominal thermal power operation level for the IEA-R1 reactor is 4.5 MW. The reduction in core size from 25 to 20 fuel elements elevates the average thermal power generation per fuel element from 187.5 kW to 225 kW or 20 \% and hence, it reduces the fuel cycle length of the reactor. Figure 14 compares the k\textsubscript{eff} as a function of irradiation time for configurations E01 and E05.

\includegraphics[width=3.85625in,height=2.61806in]{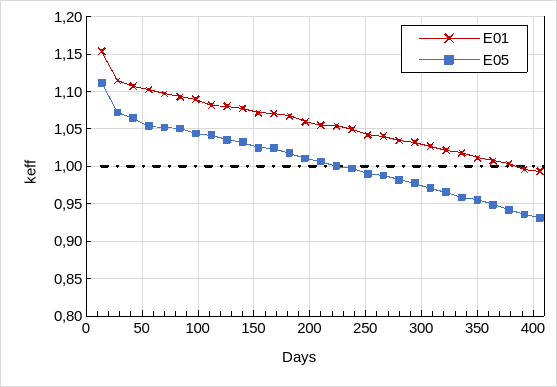}

\emph{Figure 14: Lifecycle comparison between E01 and E05 configurations. The operation thermal power level is 4.5 MW.}

Configuration E01 has a fuel cycle length of about 380 days while configuration E05, about 220 days. The fuel cycle length reduction is of 160 days or 42 \%.

\subsection{\texorpdfstring{\textsuperscript{135}Xe poisoning }{135Xe poisoning }}\label{xe-poisoning}

The xenon poisoning as a function of time for configuration E05 was evaluated and is presented in Figure 15. The final reactivity increased to around 3600 pcm due the reduction in the number of fuel elements in the core. Consequently, the thermal neutron flux increased and contributed to increase the \textsuperscript{135}Xe reactivity poisoning.

\includegraphics[width=5.39584in,height=3.04167in]{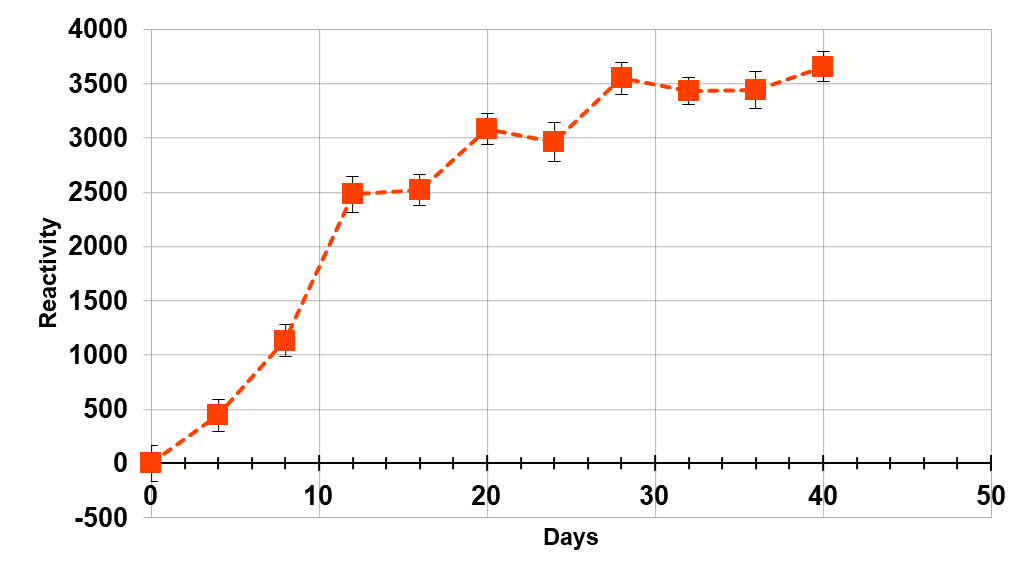}

\emph{Figure 15: Reactivity from \textsuperscript{135}Xe buildup as a function of time for the E05 configuration.}

\subsection{Lutetium-177 Production}\label{lutetium-177-production}

Figure 16 presents the production of 177Lu as a function of irradiation time for configurations E05 and E01. Considering 28 days of full-power operation, the E01 configuration produced a sample with 850.00 TBq/g, while the E05 configuration, 1025.45 TBq/g, i.e., a 20.64 \% increase.

\emph{Figure 16: Specific activity \textsuperscript{177}Lu as a function of time.}

Figure 16 results are consistent with the expected production at this flux level (\textasciitilde10\textsuperscript{14} n.cm\textsuperscript{-2}s\textsuperscript{-1}) {[}15{]}.

\subsection{Thermal analysis}\label{thermal-analysis}

As part of the evaluation process for the new core, studies in Computational Fluid Dynamics (CFD) were conducted using the operating conditions of the standard core. The goal was to confirm the applicability of the CFD model for the standard core configuration and then utilize it for the new core proposal. Table 7 presents the results of the CFD model compared with the experimental results available in the benchmark {[}16{]} for the standard IEA-R1 reactor core.

After comparing the CFD model with the experimental results of the standard configuration of the IEA-R1 reactor, the same model was applied to the new core proposal, which reduces the number of fuel assemblies to 20 and maintains the same power of 4.5MW. In this case, the cosine approach was assumed to obtain the power profile and consider the same input conditions for the temperature (32.53\ensuremath{^\circ}C) and the mass flow rate (6.27kg/s) in the instrumented fuel assembly.

\emph{Table 7: Experimental and CFD results for the IEA-R1 standard configuration.}

\begin{longtable}[]{@{}
  >{\raggedright\arraybackslash}p{(\columnwidth - 14\tabcolsep) * \real{0.1161}}
  >{\raggedright\arraybackslash}p{(\columnwidth - 14\tabcolsep) * \real{0.1343}}
  >{\raggedright\arraybackslash}p{(\columnwidth - 14\tabcolsep) * \real{0.1015}}
  >{\raggedright\arraybackslash}p{(\columnwidth - 14\tabcolsep) * \real{0.1178}}
  >{\raggedright\arraybackslash}p{(\columnwidth - 14\tabcolsep) * \real{0.1198}}
  >{\raggedright\arraybackslash}p{(\columnwidth - 14\tabcolsep) * \real{0.1161}}
  >{\raggedright\arraybackslash}p{(\columnwidth - 14\tabcolsep) * \real{0.1617}}
  >{\raggedright\arraybackslash}p{(\columnwidth - 14\tabcolsep) * \real{0.1326}}@{}}
\toprule\noalign{}
\endhead
\bottomrule\noalign{}
\endlastfoot
\multicolumn{2}{@{}>{\raggedright\arraybackslash}p{(\columnwidth - 14\tabcolsep) * \real{0.2504} + 2\tabcolsep}}{%
\multirow{2}{=}{\textbf{Steady State Conditions}}} & \multicolumn{5}{>{\raggedright\arraybackslash}p{(\columnwidth - 14\tabcolsep) * \real{0.6170} + 8\tabcolsep}}{%
\textbf{Coolant}} & \textbf{Fuel} \\
& & \multicolumn{3}{>{\raggedright\arraybackslash}p{(\columnwidth - 14\tabcolsep) * \real{0.3391} + 4\tabcolsep}}{%
\textbf{Experimental data}} & \textbf{CFD} & & \textbf{CFD} \\
\textbf{Reactor Power (MW)} & \textbf{Nominal Operating Pressure (bar)} & \textbf{Mass}

\textbf{Flow Rate (kg/s)} & \textbf{Inlet Temp.(\ensuremath{^\circ}C)} & \textbf{Average Outlet Temp.(\ensuremath{^\circ}C)} & \textbf{Average Outlet Temp.(\ensuremath{^\circ}C)} & \textbf{Relative Error (\%)} & \textbf{Maximum Temp. (\ensuremath{^\circ}C)} \\
3.5 & 1.7 & 6.27 & 32.69 & 37.57 & 37.575 & 0.01330672 & 47.107 \\
4.0 & 1.7 & 6.27 & 31.61 & 37.23 & 37.233 & 0.00805737 & 48.218 \\
4.5 & 1.7 & 6.27 & 32.53 & 38.69 & 38.697 & 0.01808926 & 50.559 \\
5.0 & 1.7 & 6.27 & 33.43 & 40.21 & 40.207 & 0.00746139 & 53.038 \\
\end{longtable}

The reactor\textquotesingle s new core underwent an evaluation process in different power density scenarios, with an average core power density of 503.28W/cm³ and a maximum power density of 710.10W/cm³. As the benchmark only reports average operating values, we compared the temperature values of the new proposal with the default core configuration, as highlighted in Table 8.

\emph{Table 8: CFD model results for the new IEA-R1 reactor core arrangement.}

\begin{longtable}[]{@{}
  >{\raggedright\arraybackslash}p{(\columnwidth - 10\tabcolsep) * \real{0.2193}}
  >{\raggedright\arraybackslash}p{(\columnwidth - 10\tabcolsep) * \real{0.1526}}
  >{\raggedright\arraybackslash}p{(\columnwidth - 10\tabcolsep) * \real{0.1484}}
  >{\raggedright\arraybackslash}p{(\columnwidth - 10\tabcolsep) * \real{0.1111}}
  >{\raggedright\arraybackslash}p{(\columnwidth - 10\tabcolsep) * \real{0.1659}}
  >{\raggedright\arraybackslash}p{(\columnwidth - 10\tabcolsep) * \real{0.2026}}@{}}
\toprule\noalign{}
\endhead
\bottomrule\noalign{}
\endlastfoot
\textbf{Scenario} & \textbf{Reactor Power (MW)} & \textbf{Inlet Temp. (\ensuremath{^\circ}C)} & \textbf{Mass Flow Rate (kg/s)} & \textbf{Outlet Temp. (\ensuremath{^\circ}C)} & \textbf{Fuel Maximum Temp. (\ensuremath{^\circ}C)} \\
\textbf{Benchmark} & 4.5 & 32.53 & 6.27 & 38.69 & - \\
\textbf{CFD of benchmark} & 4.5 & 32.53 & 6.27 & 38.697 & 50.559 \\
\textbf{CFD of E05 configuration} & 4.5 & 32.53 & 6.27 & 41.833 & 59.325 \\
\textbf{CFD of E05 configuration hottest region} & 4.5 & 32.53 & 6.27 & 45.655 & 69.323 \\
\end{longtable}

It was noted that in terms of average values, the new core does not present excessively higher values, even in its hottest region. As the standard configuration of the reference reactor has an operating range between 2 -- 5MW, we compared the results (Table 9) of the new proposal (4.5MW) with the values of the standard core for the 5MW operating scenario.

\emph{Table 9: Comparison between the new reactor core arrangement results with the 5.0MW standard operating scenario.}

\begin{longtable}[]{@{}
  >{\raggedright\arraybackslash}p{(\columnwidth - 10\tabcolsep) * \real{0.2500}}
  >{\raggedright\arraybackslash}p{(\columnwidth - 10\tabcolsep) * \real{0.1309}}
  >{\raggedright\arraybackslash}p{(\columnwidth - 10\tabcolsep) * \real{0.1651}}
  >{\raggedright\arraybackslash}p{(\columnwidth - 10\tabcolsep) * \real{0.1478}}
  >{\raggedright\arraybackslash}p{(\columnwidth - 10\tabcolsep) * \real{0.1457}}
  >{\raggedright\arraybackslash}p{(\columnwidth - 10\tabcolsep) * \real{0.1604}}@{}}
\toprule\noalign{}
\endhead
\bottomrule\noalign{}
\endlastfoot
\textbf{Scenario} & \textbf{Reactor Power (MW)} & \textbf{Inlet Temp. (\ensuremath{^\circ}C)} & \textbf{Mass Flow Rate (kg/s)} & \textbf{Outlet Temp. (\ensuremath{^\circ}C)} & \textbf{Fuel Maximum Temp. (\ensuremath{^\circ}C)} \\
Benchmark & 4.5 & 32.53 & 6.27 & 38.69 & - \\
CFD of benchmark & 4.5 & 32.53 & 6.27 & 38.697 & 50.559 \\
CFD of E05 configuration & 4.5 & 32.53 & 6.27 & 41.833 & 59.325 \\
Benchmark & 5.0 & 33.43 & 6.27 & 40.21 & - \\
CFD of benchmark & 5.0 & 33.43 & 6.27 & 40.207 & 53.038 \\
\end{longtable}

Although several approximations, the new core proposal exhibits significant potential. Consequently, a comprehensive evaluation employing specialized codes such as RELAP, PARET-ANL, MTRCR, CATHARE, and others is imperative. The assessment will provide a more detailed understanding of the proposal\textquotesingle s feasibility and efficacy, enabling to take more informed decisions regarding its implementation.

\subsection{New irradiation positions}\label{new-irradiation-positions}

Removing 4 SFE elements from the core opens the possibility of adding four more irradiation positions in the IEA-R1 core. Two additional configurations (E06 and E07) were designed to account for the irradiation potential in these new positions.

In configuration E06, irradiation elements type EIRA were inserted in the positions where the SFE were removed. In configuration E07, EIBRA type irradiation elements were inserted in the liberated positions. The positions in the core are 43, 47, 83 and 87 and refer to core matrix being 43 the upper-left position and the 87 the lower-right position. The average thermal neutron flux in the positions is around 6x10\textsuperscript{13} n/cm\textsuperscript{2}s. The EIBRA irradiation element furnishes a thermal flux about 11 \% greater than the EIRA irradiation element. The peak thermal flux in the EIBRA irradiation is greater than 8x10\textsuperscript{13} n/cm\textsuperscript{2}s in all positions and at 83 it is greater than 1x10\textsuperscript{14} n/cm\textsuperscript{2}s.

\section{CONCLUSIONS}\label{conclusions}

Changes in the core configuration of the IEA-R1 reactor were proposed aiming at producing \textsuperscript{177}Lu. The main objective was to increase the thermal neutron flux in the IEA-R1 in possible irradiation locations. Changing the core reflectors did not change much thermal neutron flux in the irradiation positions. But reducing the number of fuel elements in the core from 24 to 25 increased the average power density by 20 \% and the thermal neutron flux by between 20 to 30 \% depending on the at the irradiation positions. In addition, the configuration approximating a cylindrical arrangement demonstrated significant potential for enhancing the IEA-R1 production capacity leading to higher thermal neutron fluxes at the irradiation positions.

The evaluations included safety and operation parameters such as moderator and fuel temperature coefficients of reactivity, fuel cycle length, Xe poisoning reactivity and heat transfer and thermal analysis. All results showed that the new configurations with 20 fuel elements can be operated safely. Comparing with the current fuel burnup, the faster consumption in the E05 configuration corresponds to life cycle about 60 \% shorter. Hence, this new core configuration requires increasing fuel element production and thus to adequate the spent fuel management according to the new conditions.

Another advantage of the E05 configuration is the ability to make available four more irradiation positions. When using the EIBRA irradiation element the average neutron flux of these new location was greater than 10\textsuperscript{13} n/cm²/s with peaks at some locations of 10\textsuperscript{4} n/cm²/s. It is important to note that adding these new irradiation elements could provide 64 new irradiation locations (using EIRA elements with 2x8 locations each) or even 88 new locations (using an EIBRA with 22 locations each).

These levels of thermal neutron flux and new substantial increase in irradiation locations make the production of \textsuperscript{177}Lu in the IEA-R1 reactor viable. The changes to the reactor are relatively minor and there are essentially no new nuclear licensing issues. However, market issues and economic viability of the project must be analyzed since it will be necessary to change the reactor\textquotesingle s operating schedule, the annual demand for fuel elements will increase and more space will be needed for the storage of irradiated fuels.

\textbf{REFERENCES}

{[}1{]} IPEN. Safety Analysis Report. São Paulo: Instituto de Pesquisas Energéticas e Nucleares, 2009.

{[}2{]} BRASIL. Amendment of the Federal Constitution n. 49. {[}Constituição (1988){]}, {[}\emph{s. l.}{]}, 8 ago. 2006.

{[}3{]} SANTOS, T. A. ; Stefani, G. L. ; GENEZINI, F. A. . Optimization of IEA-R1 reactor core parameters using the particle swarm algorithm (2023). NUCLEAR ENGINEERING AND DESIGN, v. 415, p. 112713. DOI: https://doi.org/10.1016/j.nucengdes.2023.112713

{[}4{]} AHMED, Rizwan; ASLAM; AHMAD, Nasir; KHAN, Mohammad Javed; TANVIR, Mujahid. Utilization of high-density fuel and beryllium elements for the neutron flux enhancement in typical MTR type research reactors (2007). Progress in Nuclear Energy, {[}\emph{s. l.}{]}, v. 49, n. 3, p. 246--252. ISSN 01491970. DOI: https://doi.org/10.1016/j.pnucene.2006.12.002.

{[}5{]} MOKHTARI, J.; FAGHIHI, F.; KHORSANDI, J.; HADAD, K. Conceptual design study of the low power and LEU medical reactor for BNCT using in-tank fission converter to increase epithermal flux (2017). Progress in Nuclear Energy, v. 95, p. 70--77. ISSN 01491970. DOI https://doi.org/10.1016/j.pnucene.2016.11.014.

{[}6{]} SAFAEI ARSHI, S.; JOZVAZIRI, A.; MIRVAKILI, S. M.; KEYVANI, M. A methodology to enhance thermal neutron flux in Tehran Research Reactor core for domestic fuel test purposes (2021). Progress in Nuclear Energy, v. 136, p. 10372. ISSN 01491970. DOI: https://doi.org/10.1016/j.pnucene.2021.103726.

{[}7{]} STEFANI, Giovanni Laranjo; GENEZINI, Frederico Antônio; SANTOS, Thiago Augusto; MOREIRA, João Manoel De Losada. Core parametric study for enhancing the radioisotope production in the IEA-R1 research reactor (2021). Brazilian Journal of Radiation Sciences, v. 9, n. 2B, 25 jul. 2021. ISSN 2319-0612. DOI: https://doi.org/10.15392/bjrs.v9i2B.1516.

{[}8{]} SOUZA, José Antonio Batista De. Procedure for manufacturing dispersion-based fuel elements with high uranium concentration. 2011. Master in Nuclear Technology - Materials -- Universidade de São Paulo, São Paulo, 11 nov. 2011. Master's thesis DOI: https://doi.org/10.11606/D.85.2011.tde-19122011-154824.

{[}9{]} RODRIGUES, Valdemir. Monitoring report on the burn-up fuel elements in the IEA-R1 reactor. São Paulo: Instituto de Pesquisas Energéticas e Nucleares, 2022.

{[}10{]} BOSTELMANN, Rike; CELIK, Cihangir; KILE, Robert; WIESELQUIST, William. SCALE Analysis of a Fluoride Salt-Cooled High-Temperature Reactor in Support of Severe Accident Analysis (2023). 1 mar. 2022. p. ORNL/TM-2021/2273, 1854475. DOI https://doi.org/10.2172/1854475.

{[}11{]} WIESELQUIST, W. A.; LEFEBVRE, R. A.; JESSEE, M. A. SCALE Code System (2020). www.osti.gov.

{[}12{]} PILLAI, M. R. A.; CHAKRABORTY, Sudipta; DAS, Tapas; VENKATESH, Meera; RAMAMOORTHY, N. Production logistics of \textsuperscript{177}Lu for radionuclide therapy(2023). Applied Radiation and Isotopes, v. 59, n. 2--3, p. 109--118, ago. 2003. ISSN 09698043. DOI: https://doi.org/10.1016/S0969-8043(03)00158-1.

{[}13{]} SAIRANBAYEV, D.; KOLTOCHNIK, S.; SHAIMERDENOV, A.; CHAKROVA, Y.; GURIN, A.; KENZHIN, Y. Analysis of lutetium-177 production at the WWR-K research reactor (2021). Applied Radiation and Isotopes, v. 169, p. 109561. ISSN 09698043. DOI: https://doi.org/10.1016/j.apradiso.2020.109561.

{[}14{]} SILVA, Giovana Pasqualini Da. Study of the feasibility of producing Lutetium-177 in the IEA-R1 Nuclear Reactor at IPEN/CNEN-SP. 2008. Master in Nuclear Technology -- Applications (2008) -- Universidade de São Paulo, São Paulo, Master's thesis. DOI: https://doi.org/10.11606/D.85.2008.tde-11092009-174703.

{[}15{]} ZAHN, Guilherme Soares; GENEZINI, Frederico Antonio; DA SILVA, Paulo Sérgio Cardoso; NORY, Renata Mendes; MOREIRA, Edson Gonçalves; SANTIAGO, Paulo Sérgio. On the feasibility of producing Lu-177 in the IEA-R1 reactor via the direct route (2021). Brazilian Journal of Radiation Sciences v. 9, n. 1A, 30. ISSN 2319-0612. DOI: https://doi.org/10.15392/bjrs.v9i1A.1411.

{[}16{]} A. Hainoun, A. Doval, P. Umbehaun, S. Chatzidakis, N. Ghazi, S. Park, M. Mladin, A. Shokr. International benchmark study of advanced thermal hydraulic safety analysis codes against measurements on IEA-R1 research reactor (2014), Nuclear Engineering and Design, v. 280, p. 233-250, DOI: \url{https://doi.org/10.1016/j.nucengdes.2014.06.041}.

{[}17{]} D.Gong, S. Huang, G. Wang, K. Wang. Heat Transfer Calculation on Plate-Type Fuel Assembly of High Flux Research Reactor (2015), Science and Technology of Nuclear Installations, v. 2015, 13 p.DOI: \url{https://doi.org/10.1155/2015/198654}.

\end{document}